\documentstyle[sprocl,epsfig]{article}

\newcommand{\SinEff}{$\sin^2\theta^{\mbox{\scriptsize lept}}_{\mbox{\scriptsize eff}}$}
\newcommand{\sineff}{\sin^2\theta^{\mbox{\footnotesize lept}}_{\mbox{\footnotesize eff}}}
\newcommand{\mw}{$M_{\mbox{\tiny W}}$}
\newcommand{\mh}{$M_{\mbox{\tiny H}}$}

\newcommand{\mmw}{M_{\mbox{\tiny W}}}
\newcommand{\mmh}{M_{\mbox{\tiny H}}}
\newcommand{\mmz}{M_{\mbox{\tiny Z}}}
\newcommand{\mmt}{m_{\mbox{\tiny t}}}

\def\be{\begin{equation}}
\def\ee{\end{equation}}
\def\bea{\begin{eqnarray}}
\def\eea{\end{eqnarray}}

\begin{document}


\thispagestyle{empty}
\setcounter{page}{0}
\def\thefootnote{\fnsymbol{footnote}}

\begin{flushright}
IPPP/04/54\\
DESY 04-175 \\ 
DCPT/04/108 \\
SFB/CPP-04-46 \\ 
FERMILAB-Conf-04-199-T
\end{flushright}

\vspace{2cm}

\begin{center}

{\large {\bf Towards Better Constraints on the Higgs Boson Mass: \\
Two-Loop Fermionic Corrections to \SinEff }}
\footnote{talk given at International Conference on Linear Colliders LCWS~04:
19-23 April 2004, "Le Carre des Sciences", Paris, France.}

\vspace{1cm}

M.~Awramik$^{a,b}$, M.~Czakon$^{a,c}$, A.~Freitas$^d$, G.~Weiglein$^e$

\vspace*{0.8cm}

{\footnotesize
\begin{minipage}{11.5cm}
\begin{center}
$^a$ DESY, Platanenallee 6, 15738 Zeuthen, Germany\\
$^b$ Insitute of Nuclear Physics PAS, Radzikowskiego 152, 31342 Cracow, Poland\\
$^c$ Insitute of Physics, Univ. of Silesia, Uniwersytecka 4, 40007 Katowice, Poland\\
$^d$ Fermi National Accelerator Laboratory, Batavia, IL 60510-500,USA\\
$^e$ IPPP, University of Durham, Durham DH1 3LE, United Kingdom\\
\end{center}
\end{minipage}
}

\end{center}

\vspace*{2cm}

\begin{abstract}
{\normalsize
The complete two-loop electroweak  fermionic corrections
to  the  effective  leptonic  weak  mixing  angle,  \SinEff,  are  now
available.   Here   we  shortly   present  the  methods   applied  and
illustrate the implications on indirect prediction for the Higgs boson 
mass within the standard model. 
}
\end{abstract}

\newpage


\title{Towards Better Constraints on the Higgs Boson Mass: \\
Two-Loop Fermionic Corrections to \SinEff
\hspace{0.1cm}\footnote{
This work was supported in part by TMR, European Community Human
Potential Programme under contracts
HPRN-CT-2002-00311 (EURIDICE), HPRN-CT-2000-00149 (Physics at
Colliders), by Deutsche Forschungsgemeinschaft
under contract SFB/TR 9--03, and by the Polish State Committee for
Scientific Research (KBN) under contract No. 2P03B01025.
}}

\author{M.~Awramik$^{a,b}$, M.~Czakon$^{a,c}$, A.~Freitas$^d$, G.~Weiglein$^e$}

\address{$^a$ DESY, Platanenallee 6, 15738 Zeuthen, Germany\\
	$^b$ Insitute of Nuclear Physics PAS, Radzikowskiego 152, 31342
		Cracow, Poland\\
	$^c$ Insitute of Physics, Univ. of Silesia, Uniwersytecka 4, 
		40007 Katowice, Poland\\
	$^d$ Fermi National Accelerator Laboratory, Batavia, IL 60510-500,USA\\
	$^e$ IPPP, University of Durham, Durham DH1 3LE, United Kingdom\\
	}


\maketitle\abstracts{ 
The complete two-loop electroweak  fermionic corrections
to  the  effective  leptonic  weak  mixing  angle,  \SinEff,  are  now
available.   Here   we  shortly   present  the  methods   applied  and
illustrate the implications on indirect prediction for the Higgs boson 
mass within the standard model.  }

\section{Introduction}

Present efforts on both the experimental and theoretical sides
are focused on the Higgs boson. As LEPII did not give a significant 
signal of its existence, either a positive announcement at the Tevatron 
or the running of LHC are awaited. In the meantime 
the standard model is being scrutinized in order to indirectly  
predict the mass of the Higgs boson. The most valuable information comes from 
the W boson mass, \mw, and \SinEff, but also to some smaller extent
from other observables, {\it e.g.} the width of the Z boson. 
As far as $\mmw$ is concerned it is now known with one of the best
precisions, both from experiment and from theory  \cite{mw}. 
On the other hand, the weak mixing angle is 
measured with a relative precision that is almost a factor two worse  
and 
the two most accurate measurements differ from each other by $2.9\sigma$.  
Still, its dependence on the Higgs boson mass, $\mmh$,  is three times 
more pronounced than in the $\mmw$ case, demanding better theoretical 
precision for the goal of explicit tests of the model and 
for the prediction of \mh.  

Recently we performed a calculation of the two-loop fermionic
corrections to  \SinEff \hspace{0.1cm} \cite{sineff2L}.  In this
contribution we clarify some of the aspects which were not discussed
previously.   In order to avoid errors which cannot be identified
with the help of general properties of the theory (gauge invariance
and UV/IR finiteness), two independent calculations 
were performed for most parts. 
Below we sketch several details of both
methods. Finally we  demonstrate the consequences of the new result on
the Higgs boson mass prediction.
  
\section{Calculation}

The effective leptonic weak mixing angle, \SinEff, can be 
defined through the vector and axial 
vector couplings ($g_V$ and $g_A$ respectively) 
of an on-mass shell Z boson to a pair of charged leptons, such that
\be
\sineff = \frac{1}{4} \left(1-\mbox{Re} \left( \frac{g_V}{g_A}\right) \right).
\label{eq:sine}
\ee
The proper vector and axial-vector structure of the Feynman amplitude 
can be extracted by a projector. Namely, using the Dirac equation we 
find that the needed quantity is determined 
if we apply the following operator: 
\be
{\mathcal{N}}_{\mu} = \hat{p}_{2} (g_{1} \gamma_\mu + g_2 \gamma_\mu \gamma_5) \hat{p}_1
\ee  
and then take a trace over the amplitude.  Here
$\hat{p}_1$ and $\hat{p}_2$ denote momenta of the external fermions
multiplied by Dirac matrices, 
\be
g_1  =  -\frac{1}{g_A^{(0)} \; 2 \; (d-2) \; p^2},  \;\;\;\;\;
g_2  =   \frac{g_V^{(0)}}{(g_A^{(0)})^2 \; 2 \; (d-2) \; p^2},
\ee
where $p^2=(p_1+p_2)^2$, $g_V^{(0)}$ and $g_A^{(0)}$ being the tree level values of the
couplings and $d$ is the space-time dimension.
This calculation was performed in the on-shell scheme, for which 
the two-loop counter-terms were already known 
and thoroughly tested in the past \cite{muon}.   
The only complications come from the two-loop 
one-particle irreducible vertex diagrams. 
The evaluation of them was performed not only 
by two independent calculations but whenever possible also by different methods. 
\subsection{Method I }
The needed two-loop vertex diagrams may depend on  two dimensionless
variables: $\mmw/\mmz$ and/or $\mmt/\mmz$  (the Higgs boson mass does
not appear due to CP conservation).  We considered two cases: diagrams
with light fermions only, which depend on one variable; and diagrams
with top quark loops, which depend on both variables.

The light fermion contributions can be reduced to a set of master
integrals  with the help of standard methods of   Integration By Parts
and Lorenz Invariance Identities.  Still, at the two-loop level this
is a nontrivial task,  therefore it has been assigned to a newly
written C++ library, IdSolver \cite{idsolver}.
\begin{figure}[h!] 
  \epsfig{figure=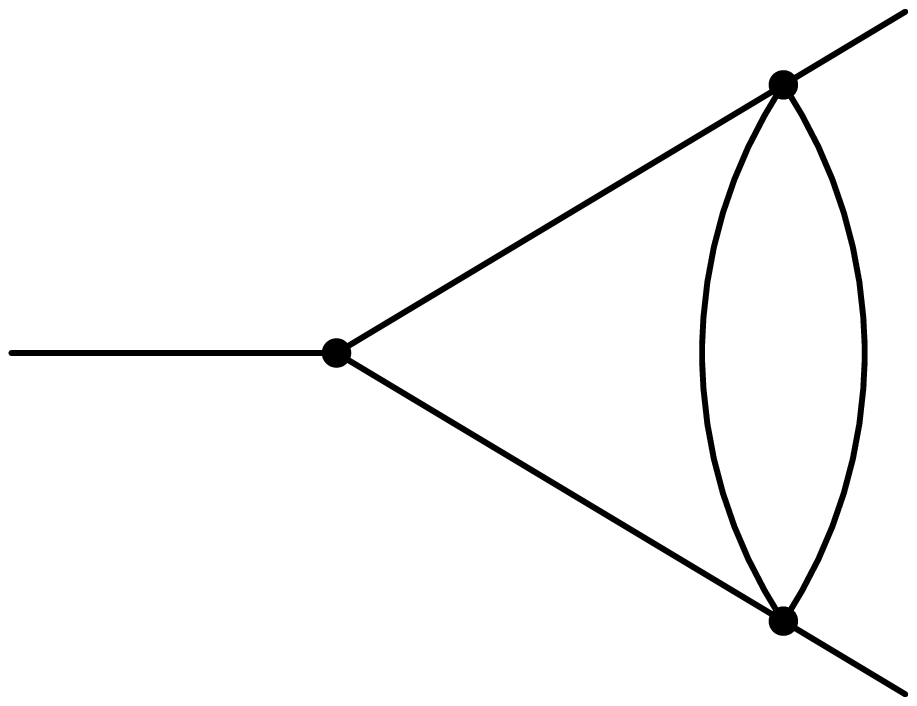, width=2.2cm}
  \epsfig{figure=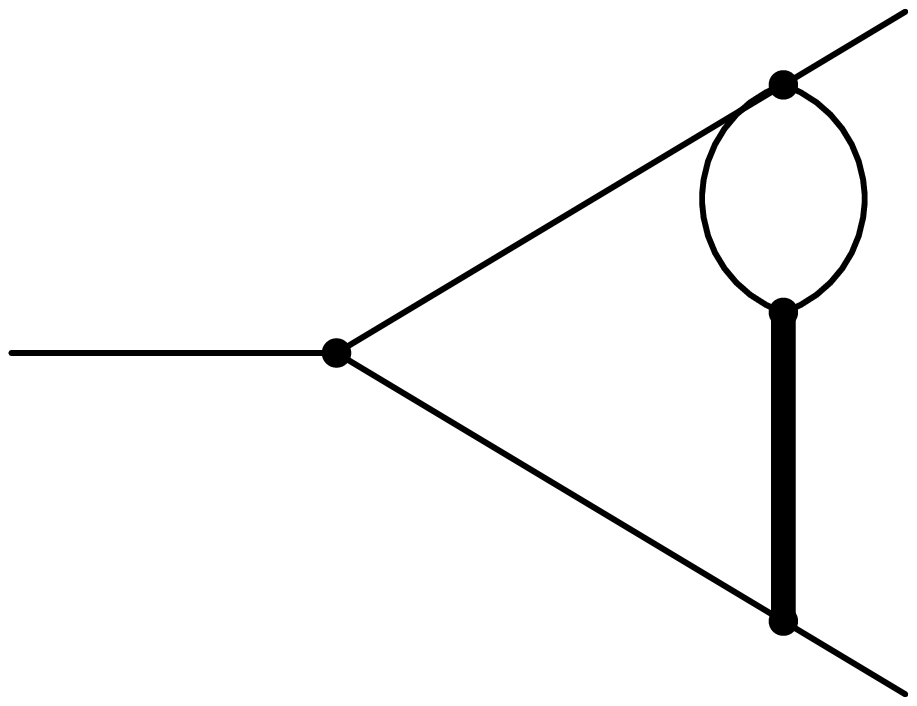, width=2.2cm}
  \epsfig{figure=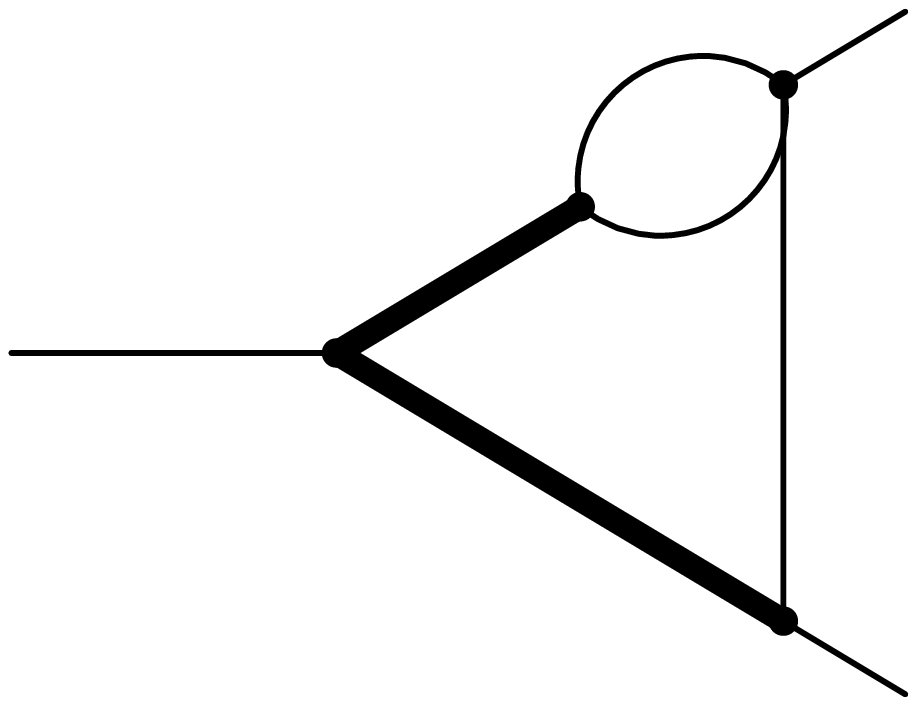, width=2.2cm}
  \epsfig{figure=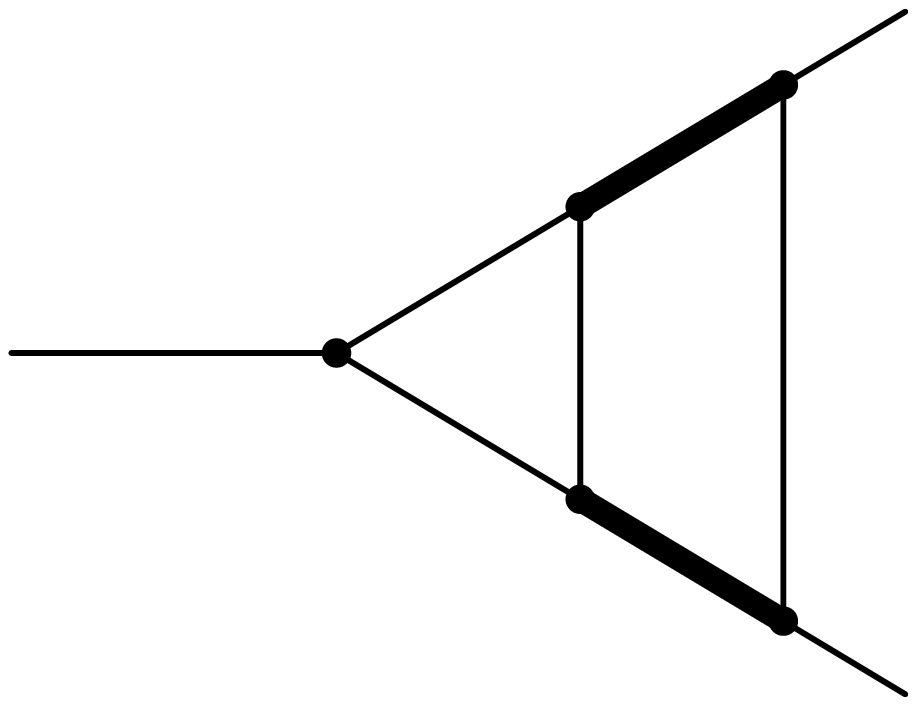, width=2.2cm}
  \epsfig{figure=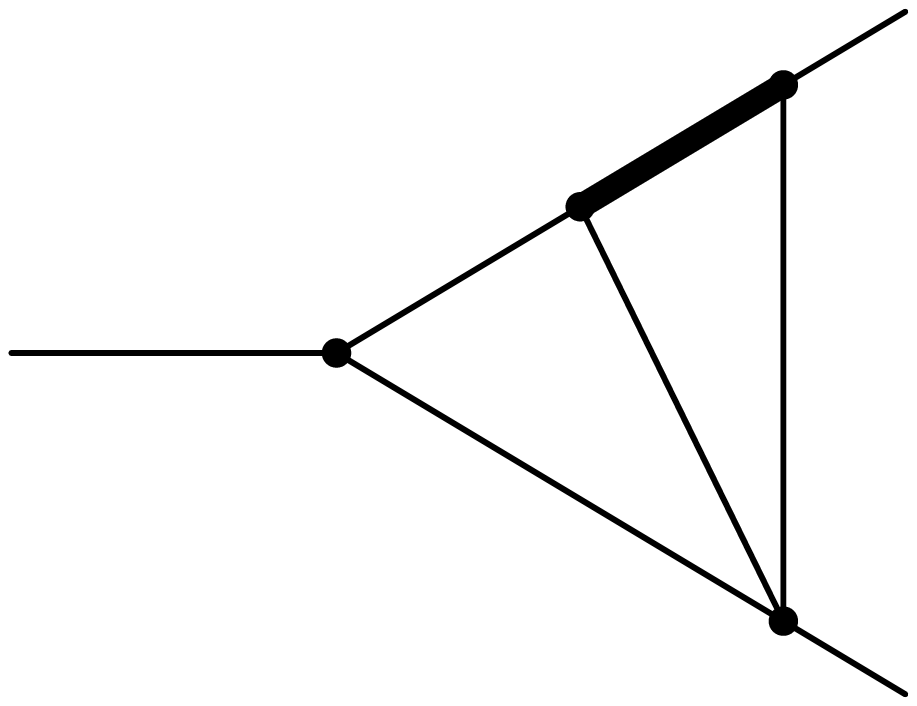, width=2.2cm}
  \caption{   
    \label{fig:masters}
    Set of master integrals required for the calculation of the light fermion contributions.
    The solid lines denote massive particles, W or Z bosons, 
    the thin lines inside a diagram denote massless particles. }
\end{figure}
The set of the identified master integrals, pictured in
Fig.~\ref{fig:masters},  was evaluated using differential equations in
the external momentum. This allowed to obtain analytical results
represented by polylogarithms of weight four at worst.

Obtaining an exact result for the heavy fermion contributions 
would be more problematic. However, 
due to large scale differences we can safely apply the heavy top mass expansion, which  
reduces the diagrams to two-loop tadpoles at most.
The series shows good convergence and in practice we used 
an expansion up to $\mmt^{-10}$, reaching sufficient precision.

For completeness we should also mention that the 
diagram generation was performed by DiaGen \cite{idsolver} and 
most of the algebraic manipulations were done with FORM \cite{form}. 
More details can be found in \cite{sineff2L}.


\subsection{Method II }

\begin{figure}[h!]
\rule{0mm}{0mm}\vspace{-1.2ex}
\begin{tabular}{c@{\hspace{1cm}}c}
\epsfig{figure=sbau_pages.epsi, width=3.7cm, bb=210 450 330 530, clip=true} &
\psfig{figure=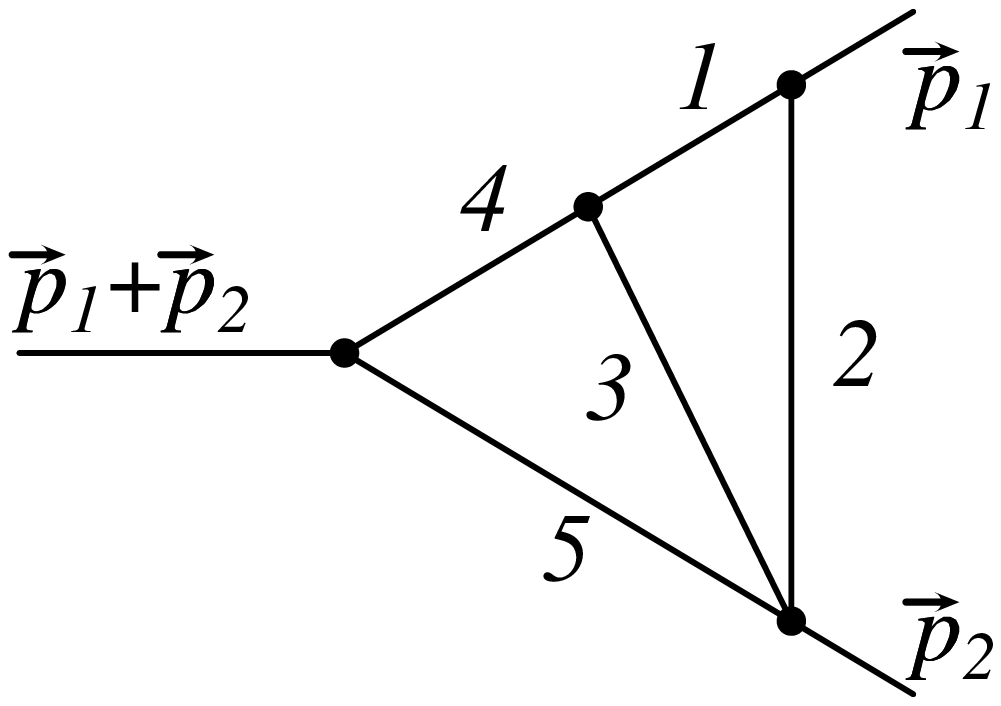, width=3cm}\raisebox{1.15cm}{$\!\!\!\to$} \
\psfig{figure=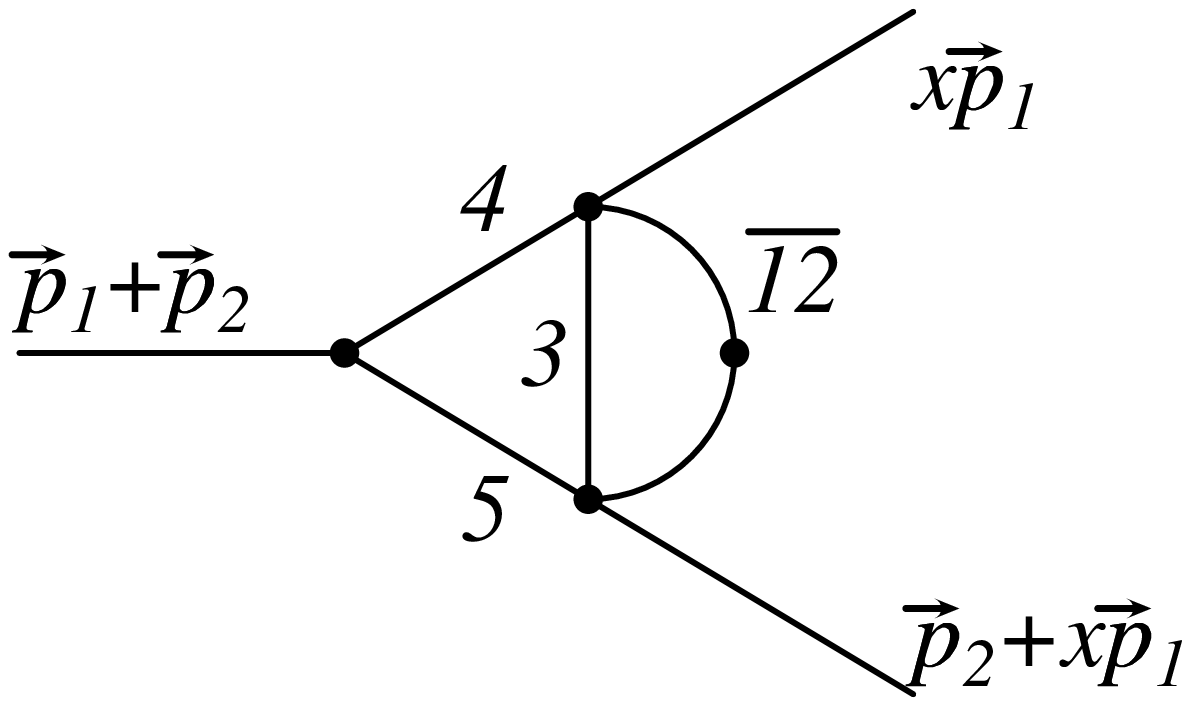, width=4cm} \\[-.5ex]
 (a) & (b) \\
\end{tabular}
\vspace{-2ex}
\caption{(a) General representation of a two-loop scalar diagram with
self-energy sub-loop. (b) Reduction of triangle sub-loop to self-energy sub-loop
by means of Feynman parameters.}
\label{fig:disp}
\end{figure}

We have also performed an independent calculation based on numerical
integrations for the master integrals. This method is based on a dispersion
representation of the one-loop self-energy function $B_0$,
\begin{eqnarray}
B_0(p^2,m_1^2,m_2^2) &=& \textstyle -\int_{(m_1+m_2)^2}^\infty {\rm d}s \,
  \frac{\Delta B_0(s,m_1^2,m_2^2)}{s - p^2}, \\
\Delta B_0(s,m_1^2,m_2^2) &=& \textstyle (4\pi\mu^2)^{4-d} \,
  \frac{\Gamma(d/2-1)}{\Gamma(d-2)} \, \frac{\lambda^{(d-3)/2}(s,m_1^2,m_2^2)}%
  {s^{d/2-1}},
\end{eqnarray}
where $\lambda(a,b,c) = (a-b-c)^2 - 4bc$.
Using this relation, any scalar two-loop integral $T$ with a self-energy
sub-loop as in Fig.~\ref{fig:disp}~(a)
can be expressed as \cite{sbau}
\begin{equation}
\begin{array}{l}
T_{N+1}(p_i;m_i^2) = - \int_{s_0}^\infty {\rm d}s \;
  \Delta B_0(s,m_N^2,m_{N+1}^2) \\[1ex]
  \times \int {\rm d}^4 q  \,
  \frac{1}{q^2-s} \,
  \frac{1}{(q+p_1)^2 - m_1^2} \cdots \frac{1}{(q+p_1+\dots+p_{N-1})^2 -
  m_{N-1}^2}.
\end{array}
\end{equation}
Here the integral in the second line is an $N$-point one-loop
function, and the integration over $s$ is performed numerically. While in
principle it is also possible to introduce dispersion relations for triangle
sub-loops, it is technically easier to reduce them to self-energy sub-loops by
introducing Feynman parameters \cite{feynpar},
\begin{equation}
\begin{array}{l}
 [(q+p_1)^2-m_1^2]^{-1} \; [(q+p_2)^2-m_2^2]^{-1} = \int_0^1 {\rm d}x
  \; [(q+\bar{p})^2 - \overline{m}^2]^{-2} \\[1ex]
\bar{p} = x\,p_1 + (1-x)p_2, \qquad
\overline{m} = x \, m_1 + (1-x) m_2 - x(1-x)(p_1-p_2)^2.
\end{array}
\end{equation}
This is indicated diagrammatically in Fig.~\ref{fig:disp}~(b). The integration
over the Feynman parameters is also performed numerically. As a result, all
master integrals for the vertex topologies can be evaluated by at most 3-dim.
numerical integrations. Similar to before, the reduction of integrals with
irreducible numerators to a small set of master integrals in the case
of propagator subloops was accomplished by using Integration By Parts
and Lorentz Invariance identities, which were implemented in an
independent realization of the Laporta algorithm \cite{laporta} 
within Mathematica. 


\section{Results and Conclusion}

\begin{figure}[t!]
\begin{center}
  \psfig{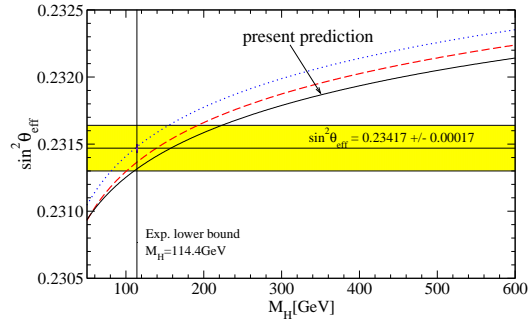}
\end{center}
  \caption{
    \label{fig:fit}
    \SinEff obtained with the previous
    and most recent fitting formulas. The blue dotted and red dashed curves denote 
    the results of $^9$
    with $\mmt=174.3$~GeV and $\mmt=178$~GeV respectively. 
    The black solid curve represents the newest 
    prediction of $^2$     
    for $\mmt=178$~GeV.     
  }
\end{figure}

Our new fitting formula, presented in  \cite{sineff2L}, contains 
all the recent results on two- and three- loop corrections to \SinEff 
(for references see \cite{sineff2L}). Using this formula, instead of
the old one \cite{mt2}, the central value of the 
Higgs boson mass is shifted by about +18.6~GeV if $\mmh$ is determined from 
\SinEff alone
\footnote{We use the same parameters as in 
\cite{sineff2L}, except for the new experimental value on \SinEff.}. 
This should be compared with an almost as large shift of +20 GeV given
by the previous formula only, which was generated by the recent change
in the measured top quark mass  \cite{newtop}.
These effects are shown in Fig.~\ref{fig:fit}. 
We do not plot the uncertainty on the theory curves; the theoretical error 
on \SinEff due to the neglect of higher order contributions is
estimated to be $4.9\times10^{-5}$ whereas the error on 
the standard model parameters, mainly $\mmt$ and $\Delta
\alpha_{had}$, gives around 
$2.6\times10^{-4}$ for $\mmh$ in the range from 100 to 600~GeV. 
In the global standard model analysis of EWWG 
(see ``blue-band plot'' in \cite{ewwg}) the net effect of these results
is not so strong setting the central value of the $\mmh$
approximately at 117~GeV with upper limit around 260~GeV (95\%CL). 

\end{document}